\documentclass[prd,english,preprintnumbers,amsmath,amssymb,nofootinbib,superscriptaddress,a4,11pt]{revtex4-1}
\usepackage[latin1]{inputenc}
\usepackage{graphicx}
\usepackage{color}
\begin{document}
\title{On unimodular quantum gravity}

\author{Astrid Eichhorn}
\affiliation{\mbox{\it Perimeter Institute for Theoretical Physics, 31 Caroline Street N, Waterloo, N2L 2Y5, Ontario, Canada}
\mbox{\it E-mail: {aeichhorn@perimeterinstitute.ca}}}

\begin{abstract} 
Unimodular gravity is classically equivalent to standard Einstein gravity, but differs when it comes to the quantum theory: The conformal factor is non-dynamical, and the gauge symmetry consists of transverse diffeomorphisms only. Furthermore, the cosmological constant is not renormalized. Thus the quantum theory is distinct from a quantization of standard Einstein gravity. Here we show that within a truncation of the full Renormalization Group flow of unimodular quantum gravity, there is a non-trivial ultraviolet-attractive fixed point, yielding a UV completion for unimodular gravity. We discuss important differences to the standard asymptotic-safety scenario for gravity, and provide further evidence for this scenario by investigating a new form of the gauge-fixing and ghost sector.
\end{abstract}

\maketitle

\section{Introduction}
In the search for an ultraviolet (UV) complete theory of quantum gravity, it is of particular interest to investigate classically equivalent formulations of gravity that differ at the quantum level. Here we explore two such options, namely standard Einstein gravity and unimodular gravity. In unimodular gravity, the determinant of the metric, $\sqrt{g}$, is not a dynamical variable, i.e.,
\begin{equation}
\sqrt{g}= \bar{\epsilon}, \label{unimodcond}
\end{equation}
where $\bar{\epsilon}$ is a constant density.
Only the conformal part of the metric is dynamical, i.e., an external notion of local scale exists.
This formulation is of particular interest for quantum gravity, as it is equivalent to General Relativity (GR) classically, see \cite{Finkelstein:2000pg}, since the equations of motion agree and each local coordinate patch of a solution to Einstein's equations admits the introduction of coordinates where $\sqrt{g} = \rm const$. At the quantum level, both theories show crucial differences, while both have only a massless spin-2 excitation as their propagating degree of freedom \cite{vanderBij:1981ym}.
An important motivation to consider unimodular gravity is that the cosmological constant arises as a constant of integration in the equations of motion and not as a coupling in the action. Thus it is not renormalized \cite{Weinberg:1988cp}. This solves the technical naturalness problem, which consists in the question why quantum fluctuations do not set the cosmological constant to the "natural" value $M_{\rm Planck}^2$. 
Note that quantum fluctuations, such as those responsible for the Lamb-shift \emph{do} gravitate in this setting, thus unimodular gravity only degravitates the cosmological constant.
Furthermore, unimodular gravity differs from GR in that the spectrum of quantum fluctuations around a background $g_{\mu \nu}$ differs, since \eqref{unimodcond} implies that $h=g^{\mu \nu} \delta g_{\mu \nu}=0$. Thus the quantum fluctuations that must be integrated over in the partition function for quantum gravity differ. This can have crucial consequences for the theory and might in particular imply, that only one of the two classically equivalent theories, GR and unimodular gravity, exists as a quantum theory.

Quantizing unimodular gravity is possible in different ways, see, e.g., \cite{Smolin:2009ti}: One possibility is to impose the  condition \eqref{unimodcond} in the action and the path-integral, thus reducing the dynamical variables in comparison to GR \cite{Unruh:1988in}. Also the gauge symmetry consists of local-volume-preserving diffeomorphisms, only. As a second option, the unimodular condition is implemented using a Lagrange multiplier \cite{Henneaux:1989zc}, thus keeping full diffeomorphism symmetry with the full metric being the variable in the path-integral. Evidently the classical equivalence of the two formulations needs not to carry over to the quantum case. Here, we focus on the first option. Possibly, the reduction in the number of unphysical degrees of freedom that are integrated over in the path-integral could reduce the regularization-scheme dependence of Renormalization Group (RG) studies of the path-integral, and thus yield more reliable results already in simpler truncations of the full space of operators.

The difference between GR and the unimodular theory becomes important already in the context of an effective-field theory setting for gravity, where quantum gravity effects are treated perturbatively, see, e.g., \cite{Alvarez:2005iy} or in semiclassical calculations, see \cite{Fiol:2008vk}. Here, we go a step further and consider a UV complete theory of GR as well as unimodular gravity. In the context of a continuum path-integral quantization of gravity, such a UV completion relies on the finiteness of running couplings in the effective action. As is well-known, a perturbative treatment reveals the perturbative non-renormalizability, not yielding a UV-complete theory, see \cite{'tHooft:1974bx,Goroff:1985sz,vandeVen:1991gw}. This is due to the negative dimensionality of the Newton coupling in $d=4$ spacetime dimensions and implies that the Gau\ss{}ian fixed point of the RG flow is ultraviolet repulsive.
In a non-perturbative context, this does not imply the breakdown of the theory in the UV, since a nontrivial fixed point of the RG flow, if found, yields a UV-completion for the effective theory, see \cite{Weinberg:1980gg}. In principle, this scenario can be realized in either standard Einstein gravity or in unimodular gravity. The first option has been studied extensively, see, e.g., \cite{Lauscher:2001ya,Litim:2003vp,Codello:2008vh,Benedetti:2009rx} following the work in \cite{Reuter:1996cp} and will be referred to as Quantum Einstein Gravity (QEG) here, for reviews see \cite{Niedermaier:2006ns,Percacci:2007sz,Litim:2008tt,Reuter:2012id}.
Here we explore the second option, Unimodular Quantum Gravity (UQG), for the first time.

We use the functional RG (FRG), where the Wetterich equation \cite{Wetterich:1993yh} allows to evaluate $\beta$ functions even in the non-perturbative regime, by using an infrared (IR) mass-like regulator $R_k(p)$ that suppresses IR modes (with $p^2 <k^2$) in the generating functional. The $k$-dependent effective action $\Gamma_k$ contains the effect of quantum fluctuations above the scale $k$ only. Its scale-dependence is given by a functional differential equation:
\begin{equation}
\partial_t \Gamma_k= \frac{1}{2} {\rm STr} \left(\Gamma_k^{(2)}+R_k \right)^{-1}\partial_t R_k.
\end{equation}
Herein $\partial_t = k\, \partial_k$, and $\Gamma_k^{(2)}$ is the second functional derivative of $\Gamma_k$ with respect to the fields and is matrix-valued in field space. Adding the regulator and taking the inverse yields the full propagator.
The supertrace contains a trace over all indices and summation over all fields with a negative sign for Grassmannian fields. The FRG framework is well-tested in diverse theories; with results agreeing with the universal one-loop $\beta$ functions of dimensionless couplings, see, e.g., \cite{Reuter:1993kw}.

We apply the background field
 formalism \cite{Abbott:1980hw}, where the metric is
split into background and fluctuation field. In contrast to the standard linear split $g_{\mu \nu}= \bar{g}_{\mu \nu}+ h_{\mu \nu}$, we have to adapt this split to the unimodular setting:
\begin{equation}
g_{\mu \nu} = \bar{g}_{\mu}^{\kappa} e^{h_{..}}_{\kappa \nu}= \bar{g}_{\mu \nu}+ h_{\mu \nu}+ \frac{1}{2} h_{\mu}^{\kappa}h_{\kappa\nu}+...\label{split}
\end{equation}
Such a non-linear split includes the unimodularity condition \eqref{unimodcond}, since, imposing $h=0$ on the fluctuations ensures that ${\rm det}g = {\rm det} \bar{g}$ to all orders in $h$. This departure from the usual linear split of the metric into background and fluctuations will also imply that the spectrum of fluctuations will change, as we will discuss below.
Note that this split does not mean that we consider only small fluctuations
around, e.g., a flat background.  Within the FRG approach we can also access the fully non-perturbative regime. The background-field formalism is used in gravity, since the
background metric  distinguishes "high-momentum" and
"low-momentum" modes by the spectrum of the background covariant
Laplacian. Background-independence follows as the $\beta$ functions do not rely on any specific field configuration, excepting possible topological subtleties \cite{Reuter:2008qx}.

Within the FRG framework as applied to quantum gravity, the unimodular theory is particularly interesting for several reasons:
Firstly, within the background field formalism, the running of background couplings receives contributions which arise from the background dependence of the regulator function. Here, this problem is reduced as the cosmological constant is not a running coupling, and therefore the scheme dependence of the RG flow is reduced.
Secondly, since the conformal factor is non-dynamical, the RG-scale $k$ becomes an external parameter, as in other quantum field theories. This is different from the setting in QEG, where $k$ can be redefined by a redefinition of the metric. Thus the meaning of large and small scales becomes dependent on an auxiliary background metric, and different choices of background correspond to a different order in which quantum fluctuations are integrated out. This differs in UQG, where an external notion of scale exists, bringing the RG flow of UQG closer to the flows of standard matter theories.
Thirdly, the conformal-factor instability of the standard Euclidean formulation is absent, as the conformal factor is not a dynamical variable in the path integral, see \cite{vanderBij:1981ym}. In QEG, the full quantum action presumably contains terms beyond an Einstein-Hilbert term which could stabilize the potential. It has been conjectured that the instability signals a non-trivial vacuum for QEG with a finite expectation value for the conformal factor \cite{Bonanno:2012dg}. The absence of this instability in UQG implies that the vacuum structure of QEG and UQG could differ.

\section{Construction of the gauge-fixing and ghost sector of UQG}
In order to construct the Wetterich equation for UQG one must introduce a gauge-fixing term, following the Faddeev-Popov trick.
The unimodularity condition \eqref{unimodcond} implies that unimodular gravity restricts the symmetry to transverse diffeomorphisms, for which
\begin{equation}
\delta_D g_{\mu \nu} = \mathcal{L}_v g_{\mu \nu} \mbox{ with } D_{\mu}v^{\mu}=0.
\end{equation}
Note that in gravity theories which are invariant under transverse diffeomorphism only, there appears an additional scalar mode in the linearized theory. As noted in \cite{Alvarez:2006uu}, this mode is absent in two cases: If the symmetry is enhanced to a full diffeomorphism symmetry, arriving at standard Einstein gravity, or if the metric determinant remains fixed. Then the additional scalar, which plays the role of the determinant, is removed from the theory.

Accordingly we cannot impose a standard linear gauge condition, such as, e.g., harmonic gauge, since it consists of four independent conditions, whereas we can impose only three if we are to respect the transversality of the diffeomorphism. One possibility is to simply project the harmonic gauge onto the transversal part, \cite{Alvarez:2008zw}, using a transverse projector $P_{\mu \nu}= \frac{1}{\bar{D}^2}\left(g_{\mu \nu}\bar{D}^2 - \bar{D}_{\mu}\bar{D}_{\nu}\right)$. Then our background-field gauge condition reads
\begin{eqnarray}
F_{\mu}&=& \sqrt{2}\left(\bar{D}^2 \bar{D}_{\kappa}h^{\kappa}_{\mu} - \bar{D}_{\mu}\bar{D}_{\rho}\bar{D}_{\sigma}h^{\rho \sigma}\right),\label{gaugecondition}
\end{eqnarray}
such that 
\begin{equation}
 S_{\rm gf}= \frac{1}{2\alpha} \int d^4x\, \bar{\epsilon} \,\bar{g}^{\mu \nu}F_{\mu}F_{\nu},\label{gfaction}
\end{equation}
where $\alpha$ is a (dimensionfull) gauge parameter, which we will later set to $\alpha=0$.
We then construct the Faddeev-Popov ghost sector as
\begin{equation}
S_{\rm gh}=- \int d^4 x \,\bar{\epsilon} \, \bar{c}_{\mu}\,\bar{g}^{\mu \nu} \frac{\partial F_{\nu}}{\partial h_{\alpha \beta}} \mathcal{L}_C g_{\alpha \beta}.
\end{equation}
Crucially, since the gauge-transformations are transverse, this implies that the ghost field also fulfills transversality: $\bar{D}_{\mu}c^{\mu}=0$. This implies that the scalar longitudinal ghost, present in QEG, is missing from the unimodular formulation. This is in complete agreement with the intuition: If the gauge-symmetry is smaller, then fewer ghost fields need to be introduced into the theory to cancel the effect of non-physical metric modes. 

Inserting the gauge condition \eqref{gaugecondition} we finally arrive at:
\begin{eqnarray}
S_{\rm gh}&=&-  \int  d^4 x \,\bar{\epsilon}\, \bar{c}_{\mu}\,\bar{g}^{\mu \nu} \Bigl( \bar{D}^2 \bar{D}^{\alpha}\bar{g}^{\beta}_{ \nu} + \bar{D}^2 \bar{D}^{\beta}\bar{g}^{\alpha}_{ \nu} \nonumber\\
&{}&\quad \quad \quad- \bar{D}_{\nu}\bar{D}^{\alpha}\bar{D}^{\beta} - \bar{D}_{\nu}\bar{D}^{\beta} \bar{D}^{\alpha} 
\Bigr) g_{\rho \beta} D_{\alpha}c^{\rho}.\label{ghaction}
\end{eqnarray}

We use a York decomposition of the fluctuation field
\begin{equation}
h_{\mu \nu}\!=\! h_{\mu \nu}^{TT}+ \bar{D}_{\!\mu}v_{\nu}+ \bar{D}_{\!\nu}v_{\mu}+ \bar{D}_{\!\mu}\bar{D}_{\!\nu}\sigma - \frac{\bar{g}_{\mu \nu}}{4} \bar{D}^2\sigma + \frac{\bar{g}_{\mu \nu}}{4}h.
\end{equation}
Here $\bar{D}^{\mu}h_{\mu \nu}^{TT}=0$, $\bar{g}^{\mu \nu} h_{\mu \nu}^{TT}=0$ and $\bar{D}^{\mu}v_{\mu}=0$. 
We redefine $v_{\mu} \rightarrow ( -\bar{D}^2 - \frac{\bar{R}}{4})^{-1/2}v_{\mu}$ and exponentiate the $\bar{g}_{\mu \nu}$-dependent Jacobians resulting from the decomposition and redefinition \cite{Lauscher:2001ya} using auxiliary Grassmann fields.

Here we will also explore a new gauge-fixing sector of QEG for the first time. Usually, a covariant gauge condition corresponding to four independent conditions is chosen. To clearly highlight the differences between UQG and QEG, we choose a different route here: Using the same gauge-fixing of the transversal diffeomorphisms in UQG and QEG, see \eqref{gaugecondition} with a dynamical density $\sqrt{g}$, we are left with one further (scalar) gauge-fixing function to choose in QEG, which we take to be $F=h$. 
Thus the additional piece of the gauge-fixing for QEG reads
\begin{equation}
S_{gf\, 2}= \frac{1}{2\alpha_2} \int d^4x\, \sqrt{\bar{g}}\,F^2,\label{scalargf}
\end{equation}
where $\alpha_2$ is a second gauge parameter.
This choice brings QEG as close to UQG as possible, since it enforces trace fluctuations in the path-integral to vanish. The crucial difference between UQG and QEG is that in UQG these fluctuations are absent from the path integral for any choice of gauge, whereas in the case of QEG we could also choose a different gauge that would keep the trace fluctuations. Note that although the fluctuations $h$ do not contribute to the RG running, they still leave an imprint in the QEG case: The gauge-fixing induces a Faddeev-Popov ghost term with scalar ghosts reading
\begin{equation}
S_{\rm scalar\, gh} = - \int d^4x\, \sqrt{\bar{g}}\, \bar{\xi}\,D^2 \xi.\label{scalargh}
\end{equation}

\section{RG flow in QEG and UQG}
Our full truncation reads
\begin{equation}
\Gamma_{k\, \rm UQG} = - 2 Z_N \bar{\kappa}^2 \int d^4x \,\bar{\epsilon}\,R + S_{\rm gf} + S_{\rm gh},
\end{equation}
in the case of UQG and with chief differences for QEG:
\begin{equation}
\Gamma_{k\, \rm QEG}\!=\! - 2 Z_N \bar{\kappa}^2\! \int\! d^4x \sqrt{g}\,R+ S_{\rm gf}+ +S_{\rm gf\, 2}+S_{\rm gh}+ S_{\rm scalar\, gh}.
\end{equation}
Herein $Z_N$ is a wave-function renormalization for the graviton, and $\bar{\kappa}^2 = \frac{1}{32 \pi G_N}$. We define a dimensionless Newton coupling $G=\frac{G_N k^2}{Z_N}$. 
In more detail, the gauge-fixing and ghost action of the two theories is given by \eqref{gfaction} and \eqref{ghaction} in the unimodular case and with the additional terms \eqref{scalargf} and \eqref{scalargh} for QEG.

There are three crucial differences between the quantum fluctuations driving the RG flow in QEG and UQG:
Firstly, the trace of the metric fluctuations is not a dynamical degree of freedom due to the unimodularity condition: Since $\sqrt{g} =\rm const $, $g^{\mu \nu} \delta g_{\mu \nu} =h =0$. Thus its quantum fluctuations are not integrated over in the path integral. In the case of QEG, the exclusion of $h$ from the path-integral is a valid choice of partial gauge, but introduces an additional scalar ghost field which is absent in UQG.
Secondly, the propagator for the graviton does not receive contributions from fluctuations in the volume factor $\sqrt{g}$: In QEG, a contribution 
\begin{equation}
(\delta^2 \sqrt{g})R \sim - h^{\mu \nu}h_{\mu \nu} R, \label{volumeflucs}
\end{equation}
 arises, which is absent for UQG, simply because $\sqrt{g}= \bar{\epsilon}$ in this case.
We will call this effect the absence of volume fluctuations. 

Thirdly, the non-linear split \eqref{split} of the metric into background and fluctuation implies that the variation of the Ricci scalar changes: This is simply due to the fact, that $\delta^2 g^{\mu \nu} = h^{\mu \kappa}h_{\kappa}^{\nu}$, whereas in the linear split, this differs by a factor of 2. The second variation of the Christoffel symbol also differs for the two cases, but this difference does not enter the second variation of the Ricci scalar, which in $d$ dimensions reads
\begin{eqnarray}
&{}&\int_x \delta^2 R \\
&=& \int_x \Bigl(\zeta h^{\mu}_{\kappa} h^{\kappa \nu}R_{\mu \nu} - h_{\mu \kappa}h^{\mu \kappa}\frac{R}{d(d-1)} + \zeta h^2 \frac{R}{d(d-1)} - h_{\mu}^{ \kappa}D_{\kappa}D_{\sigma}h^{\mu \sigma} +\frac{1}{2} h_{\mu \nu}D^2 h^{\mu \nu} + \zeta \frac{1}{2} h D^2 h\Bigr),\nonumber
\end{eqnarray}
where we have discarded total derivatives and have inserted a spherical background. For $\zeta=1$ this yields the standard variation of $R$ for the linear split $g_{\mu \nu} = \bar{g}_{\mu \nu} + h_{\mu \nu}$, whereas $\zeta =0$ yields the unimodular case, where the fluctuations are traceless and are related to the full metric through \eqref{split}.

In the following we show our results for UQG and QEG. For the RG flow of $G$, we specialize to a spherical background, where the eigenvalues of the Laplacian are known exactly \cite{Rubin:1984tc}. Note that since $\sqrt{g}=\rm const$ is a valid choice of coordinates in General Relativity, this implies that the gauge-invariant eigenvalues of the Laplacian do not differ in unimodular gravity. We can then evaluate the trace over the spectrum of fluctuations on the right-hand side of the flow equation by a direct summation over the eigenvalues of the Laplacian, employing the Euler-MacLaurin formula.
We employ a regulator of the form $R_k = \left( \Gamma_{k}^{(2)}(k^2)- \Gamma_{k}^{(2)}(-\bar{D}^2)\right)\theta(k^2 - (-\bar{D}^2))$, \cite{Litim:2001up}, which we impose directly on the eigenvalues of the covariant Laplacian, which are being summed over on the right-hand side of the flow equation. This yields the following, evidently nonperturbative result for $\beta_G = \partial_t G$:
\begin{eqnarray}
\beta_{G\, \rm UQG}&=&\!2G + G^2 \frac{3 \left(1300-309 \sqrt{13} -325 \sqrt{17} \right)}{ 936 \pi -1625 G}\label{betaUQG},\\
\beta_{G\, \rm QEG}&=&\!2 G+ G^2\frac{2 \left( 9763 -3708\sqrt{13}-975 \sqrt{17}\right)}{ 3744 \pi - 650 G}\label{betaQEG}.
\end{eqnarray}

\begin{figure}[!here]
\includegraphics[width=0.9\linewidth]{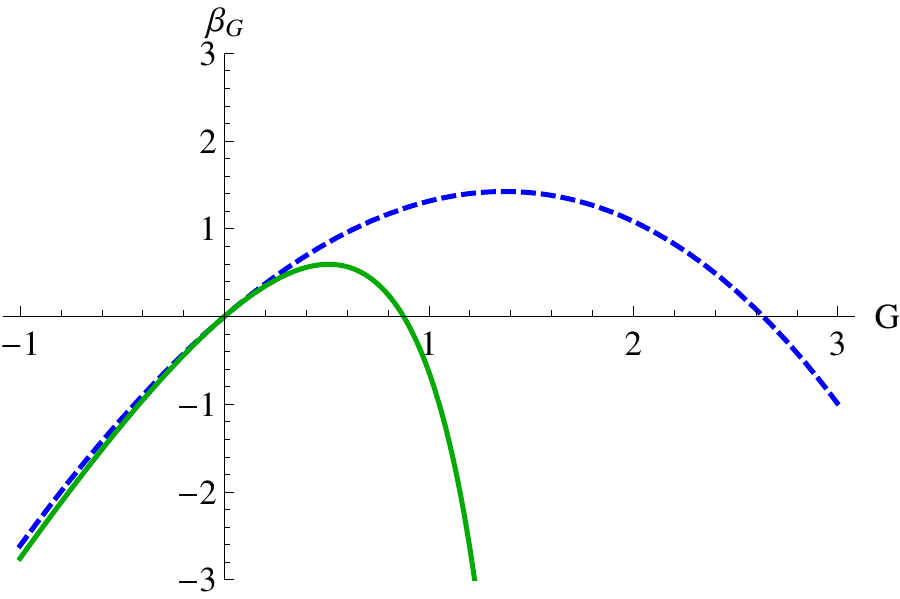}
\caption{\label{betaplot} We show the $\beta$ functions for the Newton coupling in UQG (green thick line) and QEG (blue dashed line).}
\end{figure}

Obviously, both $\beta$ functions agree in the perturbative regime, see fig.~\ref{betaplot}, as they have to due to the dimensionality of $G_N$, which stipulates $\partial_G\beta_G\vert_{G=0}=2$.

Interestingly, both $\beta$ functions show a further, interacting fixed point, which is UV attractive in both cases. Thus we find the first indication that UQG could indeed exist as a UV complete quantum theory of gravity. The critical exponent governing the approach to the non-Gau\ss{}ian fixed point at $G_{\ast\, \rm UQG}= 0.876$ in UQG, reads
\begin{equation}
\theta_{\rm UQG}=-\frac{\partial \beta_{G\,\rm UQG}}{\partial G}\Big|_{G= G_{\ast\, {\rm UQG}}}= 3.878,
\end{equation}
corresponding to a UV attractive direction.

We also find further evidence for asymptotic safety in QEG, as we find an interacting fixed point at $G_{\ast\,\rm QEG}= 2.65$ with critical exponent $-\frac{\partial \beta_{G\,\rm QEG}}{\partial G}\vert_{G= G_{\ast\rm QEG}}= 2.341$, lying in the range which is observed for this universal quantity in other schemes, see, e.g., \cite{Litim:2008tt,Codello:2008vh}.
Removing the RG improvement on the right-hand side of the flow equation, i.e., disregarding the terms $\sim \eta_N$ arising from our choice of regulator, the fixed point persists, with numerical changes in its value.

Let us discuss several approximations to the full $\beta$ function, in which the fixed point persists in both cases, which can be read as a sign of stability, and suggests that the fixed point should persist beyond our truncation. 

Firstly, we can disregard the terms $\sim \eta_N$, that arise on the right-hand-side of the flow equation due to our choice of regulator. In that case, the fixed point persists, with numerical changes in its value and the critical exponent.

Furthermore, we observe that the only mode not affected by our choice of gauge is the transverse traceless mode. In other words, a different choice of gauge would change the propagators of the other modes, and therefore their contribution to $\beta_G$, but not that of the $TT$ mode. This observation suggests to study a scenario, where all modes but the $TT$ mode are actually dropped from the flow equation, which has been dubbed the ''TT-approximation'' in \cite{Eichhorn:2010tb}. In this setting, the fixed point actually persists with only small numerical changes: $G_{\ast\, \rm UQG} = 0.93$ and $\theta_{\rm UQG} = 4.135$.\\
In the case of QEG, the changes are slightly larger, with the fixed point in the TT-approximation at $G_{\ast\, \rm QEG}= 5.41$ and a critical exponent of $\theta_{\rm QEG}= 2.85$. The fact that the change in the case of unimodular gravity is smaller could be understood from the observation that due to the reduction in the symmetry from full diffeomorphisms to transverse diffeomorphisms, a smaller portion of the fluctuation modes is actually unphysical in this case. Therefore the effect of leaving out some of these modes by going to the TT approximation is smaller in the case of UQG.

The difference between UQG and QEG relies on the off-shell character of quantum fluctuations, as the spectrum of quantum fluctuations agrees on shell: The information on the sign of the fixed point is present in the transverse traceless component of the propagator, which agrees if evaluated on shell where $R=0$, since $\Gamma_{k\, TT}^{(2)}\sim -D^2 +R \frac{3\zeta+1}{6}$, where $\zeta=1$ for QEG and $\zeta=0$ for UQG. It is actually interesting to trace the effect of the absence of volume fluctuations and the difference of \eqref{split} to the standard case: Discarding the volume fluctuations actually leads to a change of sign for the above coefficient of the $R$ term, and does indeed result in a $\beta$ function with a fixed point at $G_{\ast}<0$. Taking into account the additional difference in the spectrum of quantum fluctuations that arises due to \eqref{split} then yields a positive coefficient and a positive fixed-point value.
Note also that this analysis is in full accordance with the recent ideas of paramagnetic dominance in asymptotic safety \cite{Nink:2012vd}: The difference in the sign of the fixed-point value can be traced back to the change of sign in the curvature-dependent ("paramagnetic") term in the propagator for the transverse traceless graviton mode.

It is interesting to observe that the unimodular modes also carry information about a non-trivial fixed point in gravity, as does the conformal mode. The present study complements previous explorations of QEG in settings with reduced degrees of freedom, see, e.g., \cite{Reuter:2008wj,Eichhorn:2010tb}.

\section{Conclusions}
Here we have examined unimodular gravity and GR, which are classically equivalent, but show crucial differences in the quantum theory: Firstly, the cosmological constant is not renormalized in UQG which solves the technical naturalness problem. Secondly, the spectrum of quantum fluctuations differs in the two theories, yielding crucial differences in the RG flows. We have discussed how to set up an RG equation for the unimodular case, where the gauge-fixing and ghost sector has to be adapted to respect the unimodularity condition \eqref{unimodcond}.
We find that within a simple truncation of the RG flow both theories show an interacting fixed point for the Newton coupling and are therefore asymptotically safe. 
This is the first evidence that UQG could indeed exist as a UV complete theory. We also provide further evidence for asymptotic safety in QEG, as we investigate a new gauge fixing.

{\emph {Acknowledgements}} I thank Martin Reuter for helpful and encouraging discussions. I am also indebted to an anonymous referee for helpful comments on the implementation of the unimodularity condition.
Research at Perimeter Institute is supported by the Government of Canada through Industry Canada and by the Province of Ontario through the Ministry of Research and Innovation.

\end{document}